\documentclass[prl,twocolumn,showpacs,10pt]{revtex4-1}
\usepackage{amsmath}
\usepackage{amsfonts}
\usepackage{amssymb}
\usepackage{mathrsfs}
\usepackage{latexsym}
\usepackage{dsfont}
\usepackage{graphicx}
\newcommand{\tr}{\mbox{Tr}}

\newcommand{\bra}[1]{\ensuremath{\langle #1 |}}
\newcommand{\ket}[1]{\ensuremath{| #1 \rangle}}
\newenvironment{theorem}[1][Theorem]{\vspace{0.3cm}\noindent\textbf{#1.} }{\vspace{0.2cm}}
\newenvironment{definition}[1][Definition]{\vspace{0.3cm}\noindent\textbf{#1.} }{\vspace{0.2cm}}

\begin{document}

\title{Witness for a measure of genuine multipartite quantum discord for arbitrary N partite quantum state. }
\author{Zhi-Hao Ma}
\email[Email:]{ma9452316@gmail.com}
\affiliation{Department of Mathematics, Shanghai Jiaotong
University, Shanghai, 200240, P.R.China}

\author{Zhi-Hua Chen}
\affiliation{Department of Science, Zhijiang college, Zhejiang
University of technology, Hangzhou, 310024, P.R.China}

\date{\today }

\begin{abstract}
We propose a Witness for a  measure of genuine multipartite quantum discord for arbitrary N partite quantum state.

\end{abstract}

\pacs{03.67.-a, 75.10.Pq, 03.67.Mn}

\maketitle


Entanglement, is a distinctive feature of quantum mechanics \cite{Horodecki09,Guhne09}, and has been found numerous applications in quantum information processing tasks
\cite{Nielsen:00}.The problem of detect whether a quantum state is entanglement or not is widely studied,  However, entanglement does not necessarily exhaust all quantum correlations present in a state. Beyond entanglement, quantum discord is a suitable measure of quantum correlation.The total correlations between two quantum systems $A$ and $B$ are quantified by the quantum mutual information
\begin{equation}\label{Totale}
    \mathcal{I}(\rho_{AB})=S(\rho_{A})+S(\rho_{B})-S(\rho_{AB})
\end{equation}
where $S(\rho)=-\mathrm{Tr}(\rho \log_2 \rho)$ is the Von Neumann entropy and $\rho_{A(B)}= \mathrm{Tr}_{B(A)}(\rho_{AB})$.

On the other hand, the classical part of correlations is defined as the maximum information about one subsystem that can be obtained by performing a measurement on the other system. Given a set of projective (von Neumann) measurements described by a complete set of orthogonal projectors $\{\hat{\Pi}_{B}^{j}\}=\{|b_j\rangle\langle b_j|\}$ and locally performed only on system $B$, which satisfying that $\hat{\Pi}_{B}^{j}\geqslant0$, $\sum_k \hat{\Pi}_{B}^{j}=I$, $I$ is the identity operator, then the information about $A$ is the difference between the initial entropy of $A$ and the conditional entropy, that is $\mathcal{I}(\rho_{AB}|\{\hat{\Pi}_{B}^{j}\})=S(\rho_{A})-\sum_j p_j S(\rho_j)$, where $\rho_j=(I\otimes \hat{\Pi}_{B}^{j})\rho(I\otimes \hat{\Pi}_{B}^{j})/\mathrm{Tr}[(I\otimes\hat{\Pi}_{B}^{j})\rho(I\otimes\hat{\Pi}_{B}^{j})]$, $p_j$ is the probability of the measurement outcome $j$ and $I$ is the identity operator for subsystem $A$. Classical correlations are thus quantified by $\mathcal{Q}(\rho_{AB})=\mathrm{sup}_{\{\hat{\Pi}_{B}^{j}\}}\mathcal{I}(\rho_{AB}|\{\hat{\Pi}_{B}^{j}\})$ and the quantum discord is then defined by
\begin{equation}\label{quantum discord}
    \mathcal{D_{B}}(\rho_{AB})=\mathcal{I}(\rho_{AB})- \mathcal{Q}(\rho_{AB}),
\end{equation}
which is zero only for states with classical correlations and nonzero for states with quantum correlations. The nonclassical correlations captured by the quantum discord may be present even in separable states.\cite{Ollivier:01}

Recently, the authors get an symmetric  equivalent definition for quantum discord, the following can be found in  \cite{Rulli}:

The quantum relative entropy is a measure of distinguishability
between two arbitrary density operators $\hat{\rho}$ and $\hat{\sigma}$, which is defined as
$S\left( \hat{\rho}\parallel \hat{\sigma}\right) =\mathrm{Tr}\left( \hat{\rho}%
\log_2 \hat{\rho}-\hat{\rho}\log_2 \hat{\sigma}\right)$~\cite{Vedral:02}.
We can express the quantum mutual information $I(\hat{\rho}_{AB})$
as the relative entropy between $\hat{\rho}_{AB}$ and the product state
$\hat{\rho}_{A}\otimes\hat{\rho}_{B}$, i.e.
\begin{equation}
I\left( \hat{\rho}_{AB}\right) =S\left( \hat{\rho}_{AB}\parallel \hat{\rho}_{A}\otimes
\hat{\rho}_{B}\right).
\end{equation}
In order to express the measurement-induced quantum mutual information
$J\left( \hat{\rho}_{AB}\right)$ in terms of relative entropy, we need to consider
a non-selective von Neumann measurement on part $B$ of $\hat{\rho}_{AB}$,
which yields
$\Phi _{B}\left( \hat{\rho}_{AB}\right) = \sum_{j}
\left( \hat{1}_{A}\otimes \hat{\Pi}_{B}^{j} \right)
\hat{\rho}_{AB}
\left(\hat{1}_{A}\otimes \hat{\Pi}_{B}^{j}\right)
=\sum_{j}p_{j}\hat{\rho}_{A|j}\otimes \left\vert b_{j}\right\rangle
\left\langle b_{j}\right\vert$.
Moreover, tracing over the variables of the subsystem $A$, we obtain
$\Phi _{B}\left( \hat{\rho}_{B}\right) =
\Phi_{B}\left( \mathrm{Tr}_{A}\,\hat{\rho}_{AB}\right)
=\sum_{j}p_{j}\left\vert b_{j}\right\rangle \left\langle b_{j}\right\vert$,
where we have used that $\mathrm{Tr}_{A} (\hat{\rho}_{A|j})=1$. Then,
by expressing the entropies
$S\left( \Phi _{B}\left( \hat{\rho}_{AB}\right)\right) $ and
$S\left( \Phi _{B}\left( \hat{\rho}_{B}\right) \right) $ as
$S\left( \Phi _{B}\left( \hat{\rho}_{AB}\right) \right) = H\left( \mathbf{p}%
\right) +\sum_{j}p_{j}S\left( \hat{\rho}_{A|j}\right)$ and
$S\left( \Phi _{B}\left( \hat{\rho}_{B}\right) \right) =H\left( \mathbf{p}%
\right)$, with $H\left( \mathbf{p}\right)$ denoting the Shannon entropy
$H\left( \mathbf{p}\right) =-\sum_{j}p_{j}\log _{2}\left( p_{j}\right)$,
we can rewrite $J(\hat{\rho}_{AB})$ as
\begin{equation}
J\left( \hat{\rho}_{AB}\right) = S\left( \Phi _{B}\left( \hat{\rho}_{AB}\right) \parallel \hat{\rho}%
_{A}\otimes \Phi _{B}\left( \hat{\rho}_{B}\right) \right).
\end{equation}
Therefore, the quantum discord can be rewriten in terms of a
difference of relative entropies:
$\overline{\mathcal{D}}\left( \hat{\rho}_{AB}\right) = S\left( \hat{\rho}_{AB}\parallel \hat{\rho}%
_{A}\otimes \hat{\rho}_{B}\right) -S\left( \Phi _{B}\left( \hat{\rho}%
_{AB}\right) \parallel \hat{\rho}_{A}\otimes \Phi _{B}\left( \hat{\rho}%
_{B}\right) \right)$,
with minimization taken over $\{\hat{\Pi}_{B}^{j}\}$ to remove
the measurement-basis dependence.
It is possible then to obtain a natural symmetric extension $\mathcal{D}\left(\hat{\rho}_{AB}\right)$
for the quantum discord $\overline{\mathcal{D}}\left( \hat{\rho}_{AB}\right)$.

Indeed, performing measurements over both subsystems
$A$ and $B$, we define
\begin{eqnarray}
\mathcal{D}\left(\hat{\rho}_{AB}\right) &=& \min_{\{\hat{\Pi}_{A}^{j}\otimes\hat{\Pi}_{B}^{k}\}}
\left[  S\left( \hat{\rho}_{AB}\parallel \hat{\rho}%
_{A}\otimes \hat{\rho}_{B}\right) \right. \nonumber \\
&&\left.\hspace{-0.6cm}-S\left( \Phi _{AB}\left( \hat{\rho}%
_{AB}\right) \parallel \Phi _{A}\left( \hat{\rho}_{A}\right) \otimes \Phi
_{B}\left( \hat{\rho}_{B}\right) \right) \right] \, ,  \label{DiscordiaBipartite}
\end{eqnarray}%
where the operator $\Phi _{AB}$ is given by
\begin{equation}
\Phi _{AB}\left( \hat{\rho}_{AB}\right) =\sum_{j,k} \left(\hat{\Pi}_{A}^{j}\otimes
\hat{\Pi}_{B}^{k} \right) \hat{\rho}_{AB} \left(\hat{\Pi}_{A}^{j}\otimes \hat{\Pi}_{B}^{k}\right) \, .
\end{equation}

The aim of this work is to give a measure of genuine multipartite quantum discord for arbitrary N partite state, note that our measure is quite different from that of \cite{Rulli}.

We will extend quantum discord as given by Eq.~(\ref{DD}) to multipartite systems.

Recall that an $N$-partite pure state $|\psi\rangle\in \mathcal{H}_1\otimes \mathcal{H}_2\otimes\cdots
\mathcal{H}_N$ is called biseparable if there is a bipartition
$j_1j_2\cdots j_k|j_{k+1}\cdots j_N$ such that
\begin{equation}\label{} |\psi\rangle=|\psi_1\rangle_{j_1j_2\cdots
j_k}|\psi_2\rangle_{j_{k+1}\cdots j_N},
\end{equation}
where $\{j_1,j_2,\cdots j_{k}|j_{k+1},\cdots j_N \}$ is any
partition of $\{1,2,\cdots, N\}$, e.g., $\{13|24\}$ is a partition
of $\{1,2,3,4\}$.

Let $\gamma$ be any subset $\{j_{1}j_{2}\cdots j_{k}\}$ of $\{1, 2,..., N\}$, corresponding to
a partition $j_{1}j_{2}\cdots j_{k}|j_{k+1}\cdots j_{N}$, e.g., for three qubits state,
$\gamma=1$ corresponding to the partition $A|BC$, and corresponding to the reduced density matrix $\rho_{A}$, while if $\gamma=23$, then it corresponding to the reduced density matrix $\rho_{BC}$.

{\bf Definition.} For an arbitrary $N$ partite state $\hat{\rho}_{1 \cdots N}$, the genuine multipartite quantum discord $\mathcal{D}\left( \hat{\rho}_{1 \cdots N} \right)$ is defined as follows:

(1). First, let $\rho$ be an N partite  state, and $\gamma$ be any subset $\{j_{1}j_{2}\cdots j_{k}\}$ of $\{1, 2,..., N\}$, corresponding to
a partition $j_{1}j_{2}\cdots j_{k}|j_{k+1}\cdots j_{N}$, e.g., for three qubits state,
$\gamma=1$ corresponding to the partition $A|BC$, and corresponding to the reduced density matrix $\rho_{A}$, and $\gamma^{\prime}$ is defined as the complemental set of $\gamma$(that is, the set union of $\gamma$ and $\gamma^{\prime}$ is the total set $\{1, 2,..., N\}$, i.e.,for three qubits state, if $\gamma=1$, then $\gamma^{\prime}=23$), then define the $\gamma$-discord as

\begin{eqnarray}
\mathcal{D_{\gamma}}\left( \hat{\rho}_{1, 2,..., N}\right) &=& \min_{\{I_{\gamma^{\prime}}\otimes\hat{\Pi}_{\gamma}^{k}\}}
\left[ S\left( \hat{\rho}_{1, 2,..., N}\parallel \Phi
^{\gamma}_{1, 2,..., N}\left( \hat{\rho}_{1, 2,..., N}\right) \right) \right. \nonumber \\
&&\left.\hspace{-0.6cm} -S\left( \hat{\rho}%
_{\gamma}\parallel \Phi _{\gamma}\left( \hat{\rho}_{\gamma}\right) \right)-S\left( \hat{\rho}%
_{\gamma^{\prime}}\parallel \Phi _{\gamma^{\prime}}\left( \hat{\rho}_{\gamma^{\prime}}\right) \right)\right] .
\end{eqnarray}
where the operator $\Phi^{\gamma}_{1, 2,..., N}$ is given by
\begin{equation}
\Phi^{\gamma}_{1, 2,..., N}\left( \hat{\rho}_{1, 2,..., N}\right) =\sum_{kk^{\prime}} \left(\hat{\Pi}_{\gamma^{\prime}}^{k^{\prime}}\otimes
\hat{\Pi}_{\gamma}^{k} \right) \hat{\rho}_{1, 2,..., N} \left(\hat{\Pi}_{\gamma^{\prime}}^{k^{\prime}}\otimes \hat{\Pi}_{\gamma}^{k}\right) \, .
\end{equation}

 the superoperator $\Phi_{\gamma}$ is defined for the subsystems $\gamma$, and is given by
\begin{equation}
\Phi_{\gamma}\left( \hat{\rho}_{\gamma}\right) =\sum_{k} \left(
\hat{\Pi}_{\gamma}^{k} \right) \hat{\rho}_{\gamma} \left( \hat{\Pi}_{\gamma}^{k}\right) \, .
\end{equation}

the superoperator $\Phi_{\gamma^{\prime}}$ is defined for the subsystems $\gamma^{\prime}$, and is given by
\begin{equation}
\Phi_{\gamma^{\prime}}\left( \hat{\rho}_{\gamma^{\prime}}\right) =\sum_{k^{\prime}} \left(
\hat{\Pi}_{\gamma^{\prime}}^{k^{\prime}} \right) \hat{\rho}_{\gamma^{\prime}} \left( \hat{\Pi}_{\gamma^{\prime}}^{k^{\prime}}\right) \, .
\end{equation}

(2). then define the genuine multipartite quantum discord  as the minimal of all $\gamma$-discord:
\begin{equation}
\mathcal{D}\left( \hat{\rho}_{1, 2,..., N}\right) =\min_{\gamma}\mathcal{D_{\gamma}}\left( \hat{\rho}_{1, 2,..., N}\right)
\end{equation}

where the min run over all partition $\gamma$.

{\bf Theorem 1.} For an $N$ partite quantum state  $\hat{\rho}_{A_{1} \cdots A_{N}}$  on Hilbert space $H_{1}\otimes H_{2}\cdots H_{N}$, The genuine multipartite quantum discord $\mathcal{D}\left( \hat{\rho}_{A_{1} \cdots A_{N}} \right)$ is
non-negative, i.e., $\mathcal{D}\left( \hat{\rho}_{A_{1} \cdots A_{N}}  \right) \geqslant 0$.

Therefore, a genuine multipartite classical state can be defined by the following:
we say that a multipartite quantum state is genuine multipartite classical(GMC for short), if there exists a partition $\gamma$, such that $\hat{\rho}_{1 \cdots N} = \Phi^{\gamma}_{1, 2,..., N}\left( \hat{\rho}_{1 \cdots N} \right)$,
which means that classical states are not disturbed by a suitable local measurements.
Indeed, this definition of a classical state implies that $\hat{\rho}_{\gamma}=\Phi _{\gamma} \left( \hat{\rho}_{\gamma}\right)$, which means $\mathcal{D}\left( \hat{\rho}_{1 \cdots N} \right)=0$.

{\bf Remark.}Note that, the genuine multipartite classical (GMC)state may contain bipartite quantum discord. To see this, Define $\rho_{ABC}=\rho_{AB}\otimes\rho_{C}$,with $\rho_{AB}$ be the maximal entanglement Bell state, and $\rho_{C}=|\Phi\rangle\langle\Phi|$, $|\Phi\rangle=(1,0)^{T}$. The 3-partite state $\rho_{ABC}$ is a genuine multipartite classical state, but it contain 2-partite quantum discord.

Note that in this paper, the  local
von Neumann measurements we performed is always the form as $\{\Pi_j\} = \{\Pi_{\gamma}^{k} \otimes \Pi_{\gamma^{\prime}}^{k^{\prime}}\}$,
with  $j$ denoting the index string $(i_1 \cdots i_N)$, e.g., for 3-partite state, the measurements are: $\{\Pi_j\} = \{\Pi_{A}^{k} \otimes \Pi_{BC}^{k^{\prime}}\}$.

 Then, after a non-selective measurement,
the density operator $\rho$ becomes
\begin{equation}
\Phi(\rho) = \sum_j \Pi_j \rho \Pi_j \, .
\label{vn}
\end{equation}
This operation can then be used to define a GMC state.

\begin{definition}
If there exists any measurement $\{\Pi_j\}$ such that $\Phi(\rho)=\rho$ then $\rho$ describes a {\it genuine multipartite classical (GMC)} state
under von Neumann local measurements.
\end{definition}

Therefore, it is always possible to find out a local measurement basis such that a GMC state
$\rho$ is kept undisturbed. In this case, we will denote $\rho \in {\cal GMC}^N$, where ${\cal GMC}^N$
is the set of genuine $N$-partite classical (GMC) states. A witness for GMC states
can be directly obtained from the observation that the elements of the set $\{\Pi_j\}$ are eigenprojectors of $\rho$.
This can be shown by the theorem below (see also Ref.\cite{Luo:08,Saguia}).

\begin{theorem}
\,\,$\rho \in {\cal GMC}^N \Longleftrightarrow \left[ \rho, \Pi_j \right] = 0 \,\,\,(\forall j)$, with
$\Pi_j = \{\Pi_{\gamma}^{k} \otimes \Pi_{\gamma^{\prime}}^{k^{\prime}}\}$ and $j$ denoting the index string
$(k k^{\prime})$.
\label{t1}
\end{theorem}

We can now propose a necessary condition to be obeyed for arbitrary GMC states.

\begin{theorem}
Let $\rho$ be a GMC state, then there exists a partition $\gamma|\gamma^{\prime}$(e.g. for three partite state, $A|BC$ is such a partition), such that $\left[\rho,\rho_{\gamma}  \otimes \rho_{\gamma^{\prime}}\right] = 0$, where $\rho_{\gamma}$ be the reduced density operator for the subsystem $\gamma$, and $\rho_{\gamma^{\prime}}$ be the reduced density operator for the subsystem $\gamma^{\prime}$, e.g., $\left[\rho,\rho_{A} \otimes \rho_{BC}\right] = 0$
\end{theorem}

In \cite{Ma11}, a metric of quantum states were defined as: $D_{p}(\rho,
\sigma):=[\tr(|\rho^{\frac{1}{p}}-\sigma^{\frac{1}{p}}|^{p})]^{\frac{1}{p}}$, so $D_{2}(\rho,
\sigma):=\sqrt{2-2\tr(\rho^{\frac{1}{2}}\sigma^{\frac{1}{2}})}$.

So, if we define the following witness for state:

$W(\rho)=\min\limits_{\gamma}D_{2}(\rho,\rho_{\gamma}  \otimes \rho_{\gamma^{\prime}})$, 

This witness has deep connection with concurrence.

First, we can prove that for bipartite state, this witness is exactly the concurrence.

{\bf Theorem } For bipartite pure state $\rho:=\ket{\psi}\bra{\psi}$, $W(\rho)=D_{2}(\rho, \rho_{A}\otimes \rho_{B})=C(\rho)$, where $C(\rho)$ is the concurrence of  $\rho$, which is defined as $C(\rho):=\sqrt{2-2\tr(\rho^{2}_{A})}$.

{\bf Theorem } For multipartite pure state $\rho:=\ket{\psi}\bra{\psi}$, $W(\rho)=\min\limits_{\gamma}D_{2}(\rho,\rho_{\gamma}  \otimes \rho_{\gamma^{\prime}})=C_{GME}(\rho)$, where $C_{GME}(\rho)$ is the GME-concurrence of  $\rho$, which is defined in \cite{Ma10b} as following:

{\bf Definition } Recently, reference \cite{Ma10b} has defined a genuine multipartite entanglement measure, called GME-concurrence,  as follows:

(1). First, let $\psi$ be an n partite pure state, and $\gamma$ be any subset $\{j_{1}j_{2}\cdots j_{k}\}$ of $\{1, 2,..., N\}$, corresponding to
a partition $j_{1}j_{2}\cdots j_{k}|j_{k+1}\cdots j_{N}$, e.g., for three qubits state,
$\gamma=1$ corresponding to the partition $A|BC$, then define the $\gamma$-concurrence as
\begin{equation}C^{2}_{\gamma}(\psi)=1-\tr(\rho^{2}_{\gamma})\end{equation}

(2). Then,  define the GME-concurrence as minimal of all $\gamma$-concurrence:
\begin{equation}C^{2}_{GME}(\psi):=\min\limits_{\gamma}C_{\gamma}(\psi):=\min\limits_{\gamma}
\{1-\tr(\rho^{2}_{\gamma})\}\end{equation}

E.g., for three qubits pure state $\psi$, the GME-concurrence read as:
\begin{equation}\begin{array}{ll}C^{2}_{GME}(\psi):=\min\limits_{\gamma}C_{\gamma}(\psi):=\min\limits_{\gamma}
\{1-\tr(\rho^{2}_{\gamma})\}\\
\hspace{1.5cm}=\min\limits_{\gamma=1,2,3}
\{1-\tr(\rho^{2}_{1}),1-\tr(\rho^{2}_{2}),1-\tr(\rho^{2}_{3})\}\\
\hspace{1.5cm}=\min\limits_{A,B,C}\{1-\tr(\rho^{2}_{A}),1-\tr(\rho^{2}_{B}),1-\tr(\rho^{2}_{C})\}
\end{array}{}\end{equation}
Where $\rho_{A}$, $\rho_{B}$ and $\rho_{C}$ are the three reduced density matrices.
For  mix state $\rho$, GME-concurrence  is defined  by the convex
roof method as
\begin{equation}\label{convex}
C_{GME}(\rho):=\min\sum_{i}p_{i}C_{GME}(\psi_{i})
\end{equation}
where the minimum is taken over all decompositions of $\rho $ into pure states
$\rho =\sum_{i}p_{i} |\psi_{i}\rangle\langle\psi_{i}|$. From definition, any state $\rho$ is biseparable if and only if $C_{GME}(\rho)=0$, equivalently, the quantum state is genuine multipartite entanglement (GME) if and only if
$C_{GME}(\rho)>0$.


\end{document}